\documentclass{article}
\usepackage{graphicx      }
\title{Pulsar Wind Nebulae and Particle Acceleration 
in the Pulsar Magnetosphere}
\author{% DO NOT DELETE THIS LINE.
Shibata, S.(1)\thanks{shibata@sci.kj.yamagata-u.ac.jp},
Tomatsuri, H.(1), 
Shimanuki, M.(1), 
Saito, K.(1), \\
Nakamura, Y.(1) 
and
Mori, K.(2)
\\
\small \it
(1)Department of Physics Yamagata University, Yamagata 990-8560, JAPAN, \\
\small \it
(2)Department of Astronomy and Astrophysics, 525 Davey Laboratory, \\
\small \it
The Pennsylvania State University, University Park, PA 16802, USA, \\
\footnotesize
to appear in Proceedings of IAU 8th Asian Pacific Regional Meeting Vol. II
}% DO NOT DELETE THIS LINE.

\begin{document}
\maketitle

%%%%%%%%%%%%%%%%%%%%%%%%%%%%%%%%%%%%%
% Abstract
%
\begin{abstract}
This paper is originally intended to give a 
comprehensive review of  the pulsar wind 
nebulae and magnetosphere, but it has been 
moved to a poster paper so that we have 
changed the aim of the paper and focused on 
the Crab Nebula problem to suggest that 
particle acceleration takes place not only 
at the inner shock but also over a larger 
region in the nebula with disordered magnetic 
field. 

Kennel and Cornoniti (1984) constructed a 
spherically symmetric model of the Crab Nebula 
and concluded that the pulsar wind which 
excites the nebular is kinetic-energy dominant 
(KED) because the nebula flow induced by KED 
wind is favorable to explain the nebula spectrum
and expansion speed. 
This is true even with new Chandra observation, 
which provides newly the spatially resolved spectra. We have 
shown below with 3D modelling and the Chandra image
that pure toroidal magnetic field 
and KED wind are incompatible with the Chandra 
observation.
 
\end{abstract}

\section{Introduction}

KC model (Kennel and Coroniti 1984) assumes that a super fast MHD 
wind from the central pulsar terminates at a shock, and the shocked 
wind radiates in synchrotron radiation, which is observed as the 
nebula. The central cavity is identified as the wind region.

The pulsar wind is originally Poynting energy dominant deep in the 
pulsar magnetosphere, but by MHD acceleration the energy is converted 
into kinetic energy of the plasma outflow. How efficient the acceleration 
is a main problem in the theory of relativistic centrifugal wind. 
It is known that 
this problem is coupled with the jet-disc formation, which is clearly observed 
with Chandra (Fig. 4). 

\begin{figure}[ht]
\begin{center}
\includegraphics[width=60mm]{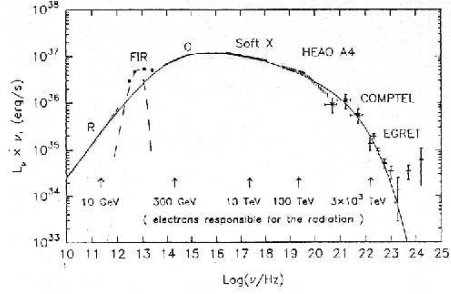}
\end{center}
\caption{KC model well expains the over all spectrum of the nebula
(after Atoyan \& Aharonian, 1996).}
\label{fig:1}
\end{figure}

The acceleration efficiency is parameterized by so-called $\sigma$-parameter
which is the ratio of the Poynting energy to kinetic energy of the 
terminal flow just before the shock. KC find that $\sigma$ determins the
expansion speed of the nebula and in turn spectrum of the nebula.
It was a great success that KC model reproduces the nebula spectrum 
(Fig. \ref{fig:1}). KC found that $\sigma = 0.003$ (kinetic energy dominant)
and the Lorentz factor of the wind is $3 \times 10^6$.

On the other hand, no wind theory explains such a high acceleration. 
It must be noted that the observed image suggest that the wind has 
jet-disc structure, which is not explained, either.

The aim of this paper is to apply the picture of KED wind to the new Chandra data
and examine whether the model still describes the nebula well or it needs
some modification.

\section{Model}

Nebula Flow is based on the KC model: \\
- toroidal magnetic field \\
- ideal MHD radial flow \\
- confined in a disc and jet (opening angles 20 degree) \\
- $L = 5 \times 10^{38}$ erg/s, $\gamma_{wind} = 3 \times 10^6$,
  $\sigma = 0.003$ \\
- power law spectrum just after the shock

\begin{figure}[ht]
\begin{center}
\includegraphics[width=60mm]{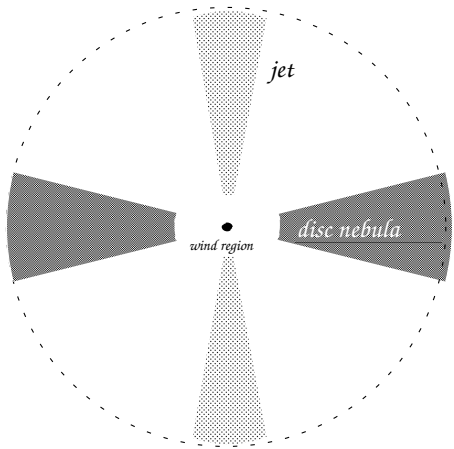}
\end{center}
\caption{Cross section of the nebula. KC flow is applied
to the disc flow region.}
\label{fig:19}
\end{figure}

\begin{figure}[ht]
\begin{center}
\includegraphics[width=60mm]{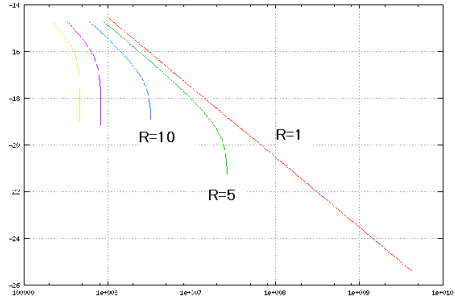}
\end{center}
\caption{Evolution of the distribution function along the flow.
We calculate the spatially resolved spetctra, which is
compared with the Chandra data (Mori 2002).}
\label{fig:2}
\end{figure}

Given flow pattern $V(R,t)$, we solve evolution of distribution function
$f(t,\epsilon)$
and magnetic field in the flow with
\begin{eqnarray}
{D  R \over Dt}             & = & V ,
\\
{D  \over Dt} \left( \ln B \right)       & = & 
- \left( {V \over R} + {\partial V \over \partial R } \right) ,
\\
{D  \over Dt} \left( \ln n \right)       & = & 
 - {D  \over Dt} \left( \ln \Gamma \right) 
 - \left( { 2 V \over R } + { \partial V \over \partial R } \right) ,
 \\
{D  \over Dt^\prime} \left( \ln \epsilon \right)       & = & 
{1 \over 3} {D  \over Dt^\prime} \left( \ln n \right) 
+ {1 \over \epsilon} \left( d \epsilon \over d t^\prime \right)_{\rm loss} ,
%\eq{fele} 
\end{eqnarray} 
and
\begin{equation}
f(t^\prime , \epsilon)= 
{n \over n_{\rm i} }
f_{\rm i} (\epsilon_{\rm i})
{d \epsilon_{\rm i} \over d \epsilon } ,
\end{equation}
where adiabatic and synchrotron losses are taken into account 
(Fig~\ref{fig:2}); $R$ is the radial distance of the fluid elecment,
$t^\prime$ is the proper time, and other notations are as usual.
Since the distribution function is obtained in the flow proper frame,
the specific emissivity is first obtained in the proper frame.
Given the observer's direction, in use of Lerentz transformation,
the emissivity is evaluated for the observer and integrated 
along the line-of-sight to obtain the observed 
brightness.

\section{Results}

\begin{enumerate}

\item 
Reproduced X-ray image is not a ring 
but `lip'-shaped (Fig.~\ref{fig:4}). This is due to the 
assumption that the magnetic field is 
pure toroidal. We suggest that some 
important fraction must be in turbulent 
field.

\item
Intensity contrast between fore and 
back sides is obtained to be 1.3,
while an observed value is about 5. 
Inconsistency is due to deceleration 
of the nebula flow, which is a result 
of small sigma value.

\item
Spatial variation of the spectra is 
well explained by the model with
small sigma (not shown here in a form of fiugre).

\item
The above points 2. and 3. are incompatible under 
a small-sigma model, suggesting
some important ingredient is missing.

\end{enumerate}

\begin{figure}[ht]
\begin{center}
\includegraphics[width=60mm]{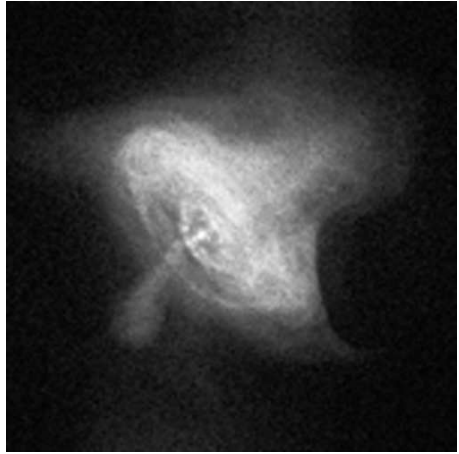}
\end{center}
\caption{Chandra image of the Crab Nebula
(Wiesskopf 2000)}
\label{fig:3}
\end{figure}
\begin{figure}[ht]
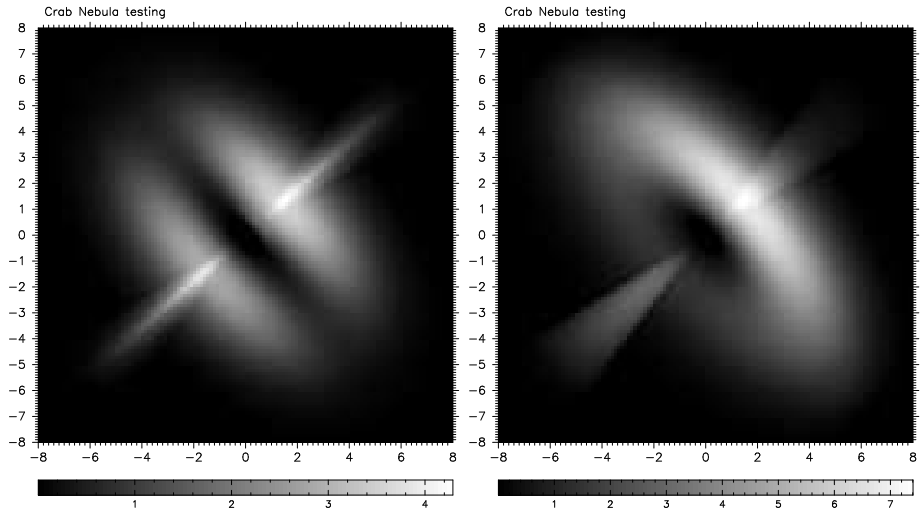

\begin{center}
\includegraphics[width=60mm]{sshibatafig5.eps}
\includegraphics[width=60mm]{sshibatafig6.eps}
\end{center}
\caption{A reproduced image of the nebula. (Left) Lip-shape is due
to the pitch angle effect with the assumption of pure toroidal field.
(Right)
If one ignore the pitch angle effect and if
$v\sim 0.2c$ (dynamically inconsistent though), then
a repurodueced image is in better agreement with the Chandra observation.}
\label{fig:4}
\end{figure}

\section{Concluding Remarks}

A simple extension of the KC model to the Chandra observation lead us to 
incompatible results: On one hand the model with a small sigma value
gives a spectrum in agreement with the observation, on the other hand 
the same model gives inconsistent intensity contrast. Together with
the suggestion of turbulent field, the nebula flow may probably be in
non-ideal MHD, which means particles are accelerated not only at the shock
but also in a larger region in the nebula by, say, magnetic reconnection.
The value of $\sigma$ is not constrained to be as small as $10^{-4}$.

%%%%%%%%%%%%%%%%%%%%%%%%%%%%%%%%%%%%%
% References
%
%\section*{References}

\begin{center} \bf references \end{center}
\begin{verse} \footnotesize
Atoyan, A. M. \& Aharonian, F. A. 1996, MNRAS 278 525 

Kennel, C. F. \& Coroniti, F. V. 1984, ApJ 283 694

Kennel, C. F. \& Coroniti, F. V. 1984, ApJ 283 710

Mori, K., 2002 PhD thesis, Ohsaka University

Wiesskopf, M. C., et al. 2000, ApJ 536 L81

\end{verse}
\end{document}